# Testing Systems of Concurrent Black-boxes—an Automata-Theoretic and Decompositional Approach*

Gaoyan Xie and Zhe Dang**

School of Electrical Engineering and Computer Science
Washington State University, Pullman, WA 99164, USA
{gxie,zdang}@eecs.wsu.edu

**Abstract.** The global testing problem studied in this paper is to seek a definite answer to whether a system of concurrent black-boxes has an observable behavior in a given finite (but could be huge) set $Bad$. We introduce a novel approach to solve the problem that does not require integration testing. Instead, in our approach, the global testing problem is reduced to testing individual black-boxes in the system one by one in some given order. Using an automata-theoretic approach, test sequences for each individual black-box are generated from the system's description as well as the test results of black-boxes prior to this black-box in the given order. In contrast to the conventional compositional/modular verification/testing approaches, our approach is essentially decompositional. Also, our technique is complete, sound, and can be carried out automatically. Our experiment results show that the total number of tests needed to solve the global testing problem is substantially small even for an extremely large $Bad$.

## 1 Introduction

Testing a concurrent and component-based system is notoriously difficult[16,14]. One difficulty comes from the system's nondeterminism and the synchronizations among concurrently running components. Another difficulty lies in the fact that, in a component-based system, its constituent components could be some externally obtained software components (such as COTS products) whose source codes and design details are usually not available. In that case, traditional white-box techniques (like static analysis) are not applicable to analyzing the system. These components can be readily treated as *black-boxes* whose models (both at code level and design level) are unknown. In this paper, we study a testing problem for such a system of concurrent black-boxes.

In our setup, a system of concurrent black-boxes consists of a host system (called the gluer) and a number of black-boxes. Each of the gluer and the black-boxes is called a *unit* (or a component), which is a (possibly nondeterministic and infinite-state) labeled transition system, each of whose labels represents either an observable action or an internal action. All the units in the system run concurrently and synchronize on a number of observable actions. The gluer is a fully specified finite-state unit. For each black-box, however, except for its interface (i.e., the set of its observable actions), everything else is unknown, while its implementation is always available and can be black-box tested. A *global bad behavior* is an observable behavior of the system in a given finite set $Bad$. Finally, the *global testing problem* studied in this paper is to verify (with a definite answer) that, for the given set $Bad$, the system does not have a global bad behavior.

A straightforward approach to solve the global testing problem is to perform integration testing over the system as a whole and see if the system exhibits a bad behavior. However, there are fundamental difficulties with this approach. For instance, in some applications [30], integration testing may not be applicable at all. Even when integration testing is possible in some situations, the system itself is often nondeterministic. The combinatorial blow-up on the number of the executions caused by nondeterministic interleavings among the concurrent units in the system generally makes it infeasible to do thorough integration testing, while we are looking for a definite answer to the global testing problem. Due to the same reason, even when one has a way to handle the nondeterminism [31], the size of the given set $Bad$ (which could be very large, e.g., more than $10^{24}$ in some of our experiments shown later) may also make exhaustive integration testing infeasible.

* The research was supported in part by NSF Grant CCF-0430531.
** Corresponding author (zdang@eecs.wsu.edu).

A less straightforward approach is to combine testing with some formal method. For instance, one can extensively test each black-box alone and try to build [26] a partial model of the black-box from the test results. Then, one can run a formal method like model-checking on the partial system model built from the partial models of the black-boxes to solve the global testing problem. However, this approach is also difficult to implement. For instance, it is hard to choose effective test sequences to build a partial model of a black-box, and it is also hard to know when the tests over a black-box are adequate. Moreover, the partial (and hence approximated) system model might not help us obtain a definite answer to the global testing problem. To avoid the above difficulties, one may also try, using some formal method, to derive an expectation condition over a black-box's behaviors such that: when every black-box behaves as expected, the system guarantees to not have a global bad behavior. Then the expectation conditions can be used to generate test sequence for the black-boxes. However, the interactions among the concurrent black-boxes make it difficult to derive such conditions automatically (see Section 2 for related work on the assume-guarantee style reasoning).

In this paper, we introduce a novel approach (called the "push-in" technique) to solve the problem, which does not entail any integration testing. Instead, in our approach, the global testing problem is reduced to testing individual black-boxes in the system one by one in some given order. Using an automata-theoretic approach, test sequences for each individual black-box are generated from the system's description as well as the test results of black-boxes prior to the black-box in the given order. Suppose that $B_1, \ldots, B_k$ represent the concurrent black-boxes in a system. The first step of our approach is to compute an auxiliary set $\mathcal{A}_1$ of sequences of observable actions for black-boxes $B_1, \ldots, B_k$ and a set $\mathcal{U}_1$ of test sequences for black-box $B_1$. Then we test the black-box $B_1$ with test sequences in $\mathcal{U}_1$ and collect all successful test sequences into a surviving set $SUV_1$. In the second step, from the surviving set $SUV_1$ and the auxiliary set $\mathcal{A}_1$, we compute the auxiliary set $\mathcal{A}_2$ (for black-boxes $B_2, \ldots, B_k$) and the test sequence set $\mathcal{U}_2$ for black-box $B_2$. Again, after testing black-box $B_2$ with test sequences in $\mathcal{U}_2$, we collect all successful testing sequences into a surviving set $SUV_2$. Subsequent steps follow similarly, and eventually, in the last step (i.e., step $k$), the global testing problem will be decided from the surviving sets. That is, the system has no global bad behavior iff, for some $1 \leq i \leq k$, the surviving set $SUV_i$ is empty. We also provide a procedure to recover a global bad behavior when the answer to the original problem is "no".

Since the sets (i.e., $\mathcal{U}_i$ and $\mathcal{A}_i$) are provably finite and, in many cases, huge, we use (finite) automata that accept the sets as their symbolic representations, and standard automata operations are used to manipulate these sets. Also, the global testing problem is decomposed into a series of testing problems over each individual black-box in the system. Hence, our approach is an automata-theoretic and decompositional approach. Moreover, the "push-in" technique is both complete and sound, and can be carried out automatically. In particular, we show that the technique is "optimal" in the sense that each test we run over a black-box has the potential to discover a global bad behavior (i.e., we never run useless tests). In general, exhaustive integration testing over a concurrent system is infeasible. However, our experiments show that, using the push-in technique, we can completely solve the global testing problem with a substantially smaller number of tests over the individual black-boxes, even for an extremely large set of $Bad$ (some of our experiments performed only about $10^5$ unit tests for a $Bad$ of size more than $10^{24}$).

The rest of this paper is organized as follows. In Section 2, previous work related to this paper is discussed. In Section 3, the formal definitions for a system of concurrent black-boxes and its global testing problem are presented. In Section 4, the detail of the push-in technique is shown. In Section 5, a set of experiments are run and the results are analyzed. Finally, Section 6 points out some future work.

## 2 Related Work

The global testing problem is essentially a verification problem since we are looking for a definite answer. In the area of formal verification, there has been a long history of research on exploiting compositionality in system verification, and a common technique is to follow the "assume-guarantee" reasoning paradigm [21,28,19,7,2,9,8,3]. However, a successful application of the paradigm depends on the correct assumptions for the components in a system, which are, in general, formulated manually. Several authors suggest solutions to the problem of automated assumption generation [17,18,12,15]. But the solutions require that the source code and/or the finite-state design is available for a unit, which, unfortunately, is not the case in our setup. Although our push-in technique relies on black-box testing instead of an "assume-guarantee" mechanism, it can be extended to a system where a black-box is associated with environmental assumptions.

In the area of software testing, researchers have long recognized the importance of combining formal methods (like model-checking) and testing techniques for system verification. Most work (e.g., [6,10,13]) stems from the spirit of specification-based testing, and utilizes model-checkers' capabilities of generating counter-examples from a system's specification to produce test-cases against an implementation. This approach typically works at the unit level and lacks a "control" over the generated test-cases, since, unlike our technique, it does not have an overall and analytical characterization over all the useful (i.e., has the potential to recover a global bad behavior) test sequences. In contrast to our ideas, theoretical work in [26,35] focuses on complete testing over a *single and finite-state* black-box with respect to a temporal property. The decompositional approaches proposed in [11,22] for model-checking feature-oriented software designs rely totally on model-checking techniques (no testing) and could cause false negatives. Integration testing of concurrent programs in [31,20] relies on a specification (unavailable in our model) of a concurrent program.

The quality assurance problem for component-based software has attracted lots of attention in software engineering. However, most work considers the problem from component developers' point of view; i.e., how to ensure the quality of components before they are released (e.g., [25,34,33,29]). This view, however, is fundamentally insufficient: an extensively tested component (by the vendor) may still not perform as expected in a specific deployment environment, since the deployment environments of a component could be quite different and diverse such that they may not be thoroughly tried by the vendor. Our push-in technique approaches this problem from system developers' point of view: how to ensure that multiple components function correctly in a host system where the components are deployed. In our technique, test sequences run on a component are customized to its specific deployment environment. Unlike our approach, frameworks like [4] require a complete specification about the component to be incorporated into a system, which is not always possible.

## 3 Preliminaries

In this paper, we consider a system of (concurrent) black-boxes, which consists of a host system (called the *gluer*) and a collection of black-box components (simply called *black-boxes*). Each of the gluer and the black-boxes is a *unit*. In the rest of the section, we will present the model of a unit, the model of the system of black-boxes, and the global testing problem for the system.

### 3.1 The Unit Model

A unit is a nondeterministic and labeled *transition system* $T$ that moves from one state to another while performing an action. Formally, $T = \langle S, s_{\text{init}}, \nabla, R \rangle$, where $S$ is an (infinite and countable) set of states with $s_{\text{init}} \in S$ being the *initial* state, $\nabla$ is a finite set of actions, and $R \subseteq S \times \nabla \times S$ defines the transition relation. In particular, the action set $\nabla$ is partitioned into three disjoint subsets: $\{\epsilon\}$ (an internal action), $\Pi$ (input actions), and $\Gamma$ (output actions). Especially, the set $\Sigma = \Pi \cup \Gamma$, i.e., the set of *observable actions* in $T$, is called the *interface* of $T$. When the set $S$ of states is a finite set, $T$ is called a *finite-state transition system*.

A *behavior* of $T$ is a sequence of actions in $\nabla$: $a_1 \ldots a_h$ (for some $h$) such that there is a sequence of states $s_0 \ldots s_h$ with $s_0 = s_{\text{init}}$ and $(s_j, a_j, s_{j+1}) \in R$ for each $0 \leq j \leq h-1$. An *observable* behavior of $T$ is the result of dropping all the internal actions (i.e., $\epsilon$'s) from a behavior. Trivially, the empty string is an observable behavior for any unit $T$.

A (unit) test sequence $\alpha$ for $T$ is a sequence of observable actions in $\Sigma$. A unit $T$ is considered to be a *black-box* if its interface (i.e., $\Pi$ and $\Gamma$) is the only known part in its definition. In this case, we assume that $T$ is testable. That is, there is a black-box testing procedure $\mathbf{BBtest}(T, \cdot)$ [1] such that, for any test sequence $\alpha$, $\mathbf{BBtest}(T, \alpha)$ returns "yes" (i.e., $\alpha$ is *successful*) if $\alpha$ is an observable behavior of the unit $T$, and, $\mathbf{BBtest}(T, \alpha)$ returns "no" (i.e., $\alpha$ is *unsuccessful*) if otherwise.

For example, consider the black-box Comm in Figure 1, which has seven observable actions (in the figure, we use suffixes ? and ! to distinguish input and output actions respectively). Assume that the black-box is implemented as shown in Figure 5. Clearly, $send\ msg\ ack$ is a successful test sequence to Comm while $send\ msg\ fail$ is not.

Obviously, if one further assumes that the black-box is output deterministic (i.e., an input action sequence uniquely decides the corresponding output action sequence), then a test sequence for the black-box can be simply reduced to a sequence of input actions. However, there are testable units that are not necessarily output deterministic (e.g., [24,32,27]). Therefore, to make our algorithms (presented later) more general, we do not apply this assumption (under which, obviously, our algorithm still applies). That's why in our definition, a test sequence is always a sequence of both input actions and output actions.

### 3.2 The System Model

A system of concurrent black-boxes consists of a gluer $G$ and a number of black-boxes $B_1, \ldots, B_k$, written $Sys = G(B_1, \ldots, B_k)$. The gluer and the black-boxes are all units which run concurrently and synchronize on certain actions. More precisely, $G$ is a fully specified and (nondeterministic) finite-state unit $G = \langle S_0, s^0_{\text{init}}, \nabla_0, R_0 \rangle$, whose interface is $\Sigma_0 = \Pi_0 \cup \Gamma_0$. Each $B_i$ is a black-box unit $B = \langle S_i, s^i_{\text{init}}, \nabla_i, R_i \rangle$, which is testable and whose interface (the only given part of the black-box) is $\Sigma_i = \Pi_i \cup \Gamma_i$. As mentioned earlier, a black-box is not necessarily a finite-state unit. The state sets $S_0, \ldots, S_k$ are all disjoint. But the interfaces $\Sigma_0, \ldots, \Sigma_k$ may not be disjoint: some units may share some common actions.

We use $\Sigma = \Sigma_0 \cup \ldots \cup \Sigma_k$ to denote all the observable actions in the system $Sys$ (this implies that each unit's observable actions are also observable in the system), and use $Sig(a)$, called the *signature* of $a$, to denote the set of all $0 \leq i \leq k$ such that $a \in \Sigma_i$. Therefore, the signature indicates the units that share action $a$.

The system $Sys$, which also works as a labeled transition system, is a Cartesian product of its units. That is, $Sys = \langle S, s_{\text{init}}, \nabla, \mathbf{R} \rangle$, where $S = S_0 \times \ldots \times S_k$ is the system's (global) state set $S$; each unit starts from its own initial state; i.e., the initial global state $s_{\text{init}}$ of the system is $(s^0_{\text{init}}, \ldots s^k_{\text{init}})$; and $\nabla = \{\epsilon\} \cup \Sigma$ with $\Sigma = \Sigma_0 \cup \ldots \cup \Sigma_k$ is the system's action set.

---

[1] The black-box testing procedure can be implemented in practice for a variety of transition systems [5].

The system's (global) transition relation $\mathbf{R} \subseteq S \times \nabla \times S$ is more complex. A global transition that moves the system from a global state $(s_0, \ldots, s_k)$ to another global state $(s'_0, \ldots, s'_k)$ while performing an action $a \in \nabla$ is in $\mathbf{R}$ iff one of the following conditions is satisfied:

- $a$ is an internal action (i.e., $\epsilon$), and exactly one unit in the system performs the internal action while the remaining units do not move; i.e., $\exists 0 \leq i \leq k.\ (s_i, \epsilon, s'_i) \in R_i \wedge \forall 0 \leq j \neq i \leq k.\ s_j = s'_j$,
- $a$ is an observable action (i.e., $a \in \Sigma$), and all the units whose interfaces contain the observable action $a$ synchronize over the action while the remaining units do not move; i.e., $\forall 0 \leq i \leq k.\ (i \in Sig(a) \wedge (s_i, a, s'_i) \in R_i) \vee (i \notin Sig(a) \wedge s_i = s'_i)$.

In other words, at any moment in the system $Sys$, exactly one unit performs an internal action, exactly one unit performs an observable action that is not shared with any other unit, or multiple units synchronize over a common observable action. It shall be noticed from the above definition that the synchronizations allowed in our model are quite flexible. Not only can the units in a system synchronize over an output/input pair as most other system models allow, they can also synchronize over just an output action or an input action, if only they can perform this (no matter output or input) action at a certain global state. Also, in our model, a synchronization can either occur between a pair of units or among more than two units; thus multi-cast or broadcast is allowed. Certainly in some systems, multi-cast, broadcast, or synchronizations over only an output action or input action may be undesirable. In that case, they can be easily eliminated just by renaming the actions. It shall also be pointed out that, in the system $Sys$, if a global transition is a synchronization over a pair of output and input actions among some units, these two actions are considered to be one single action, and we do not discriminate whether it is output or input but just treat it as an observable action to the environment.

As defined earlier, a sequence $\alpha \in \Sigma^*$ is an observable behavior of the system $Sys$ of black-boxes if the system, treated as a transition system, has an execution from the initial global state to some global state and, on the execution, $\alpha$ is the observable behavior.

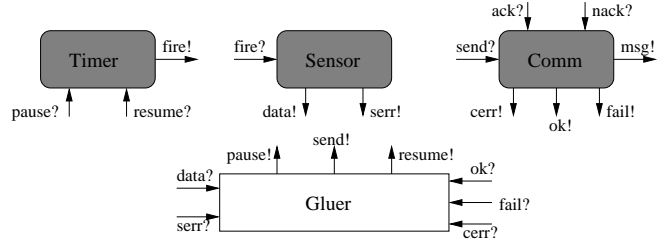

**Fig. 1.** A Data Acquisition System

For example, consider a data acquisition system shown in Figure 1, which consists of one `Gluer` and three black-box components: `Timer`, `Sensor` and `Comm`. The system works as follows. Once started, the `Timer` keeps signaling a *fire* event when the time interval set runs out; the `Timer` can also be paused (resp. resumed) by an incoming *pause* (resp. *resume*) event. The `Sensor` is supposed to respond to a *fire* event by signaling a *data* event when the sensor's reading is ready; it also signals a *serr* event when something is wrong inside the `Sensor`. The `Comm` component responds to a *send* event to send some data by signaling a *msg* event to some underlying network; it responds to an *ack* (resp. *nack*) event by signaling an *ok* (resp. *fail*) event to indicate that the data associated with a previous *send* event has been transmitted successfully (resp. unsuccessfully) by the underlying network; it signals an *cerr* event when something is wrong inside Comm. The Gluer (whose transition graph is depicted in Figure 2) simply relays data from Sensor to Comm; it pauses the Timer when something is wrong with the Sensor or Comm, and after that, it resumes the Timer when either an *ok* or *fail* is received from Comm. Together, they constitute a data acquisition system, which periodically transmits a reading of the Sensor through Comm via some underlying communication network. In this system, the Gluer and the three components run concurrently and synchronize with each other by sending and receiving those events (here, all synchronizations are over output/input pairs between two units). The internal implementations of the three components are shown in Figure 3, Figure 4, and Figure 5, respectively [2]. It can be seen (though not obviously) that the following sequence is an observable behavior of the system: *fire fire serr pause data send msg ack ok resume fire*, while sequence *fire fire serr data pause send* is not.

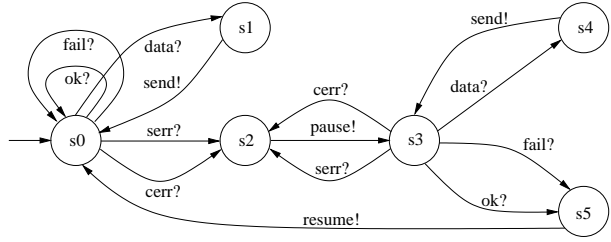

**Fig. 2.** The Gluer

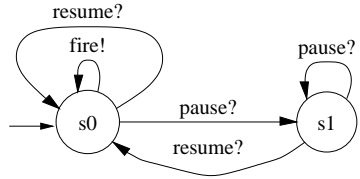

**Fig. 3.** Internal implementation of Timer

When all the black-boxes are fully specified, our system model is roughly equivalent to the IOTS studied in [27]. Our model is also closely related to I/O automata [23] (but ours is not input-enabled) and to interface-automata [9] (but ours, similar to the IOTS, makes synchronizations between units observable at the system level). These observable synchronizations are the key to testing the behavior of a system of concurrent black-boxes, where an abstract model (such as design or source code) of each black-box is unavailable.

Let $Bad \subseteq \Sigma^*$ be a given set of test sequences that are not supposed to be the observable behaviors of the system $Sys$. The *global testing problem* is to verify (with a definite answer) that none of the test sequences in $Bad$ is an observable behavior of the system. Clearly, in general, the problem can not be solved completely since the set $Bad$ can be infinite and, for testing, only finitely many test sequences can be run. Therefore, we assume that $Bad$ is a finite set, which can be given as an explicit list of test sequences (e.g., $Bad = \{fire\ fire, fire\ fire\ data, fire\ data\ send\ fire\}$) or as a symbolic representation (e.g., $Bad$ is all sequences in regular expression $fire\ data\ (fire)^*\ send$ whose lengths are between 10 and 30).

---

[2] Obviously, the push-in technique does not require these transition graphs, which are provided only for readers to understand the system

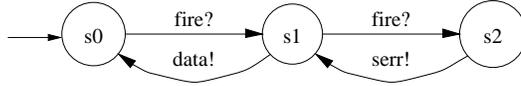

**Fig. 4.** Internal implementation of Sensor

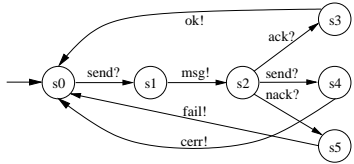

**Fig. 5.** Internal implementation of Comm

## 4 The Push-in Technique

In this section, we present the "push-in" technique to completely solve the global testing problem, by performing unit testing over each individual black-box in the system. A test sequence is a string or a word. A finite set of test sequences is therefore a regular language and, in this paper, we use a (finite) automaton that accepts the finite set as the symbolic representation of the set. Our push-in technique is automata-theoretic. For each $1 \leq i \leq k$, the technique generates two automata: $U_i$ and $A_i$. Automaton $U_i$, called a *unit test sequence automaton*, accepts words in alphabet $\Sigma_i$; i.e., it represents a set of test sequences for black-box $B_i$. Automaton $A_i$, called an *auxiliary automaton*, accepts words in alphabet $\Sigma_i \cup \ldots \cup \Sigma_k$ (observable actions for the black-boxes $B_i, \ldots, B_k$). Our push-in technique works in the following $k$ steps, where $i$ is from 1 to $k$:

**Step** $i$. The step consists of two tasks:
(Automaton Generation) This task generates the unit test sequence automaton $U_i$ and the auxiliary automaton $A_i$. We first generate the auxiliary automaton $A_i$. Initially when $i = 1$, the generation is based on the $Sys$'s description (i.e., the gluer $G$ and the interfaces for $B_1, \ldots, B_k$) and the given set $Bad$. When $i > 1$, the generation is based on the auxiliary automaton $A_{i-1}$ and the surviving set $SUV_{i-1}$ (see below) obtained from the previous **Step** $i - 1$. If the empty string is accepted by the auxiliary automaton $A_i$, then the global testing problem (none of observable behaviors of the system $Sys$ is in $Bad$) returns "no" (i.e., a bad behavior of the system exists) – no further steps need to run. We then generate the unit test sequence automaton $U_i$ directly from the auxiliary automaton $A_i$ constructed earlier. This task is purely automata-theoretic and does not involve any testing.
(Surviving Set Generation) In this second task, using **BBtest**, we perform unit testing over the black-box $B_i$ for all test sequences accepted by the test sequence automaton $U_i$ ($U_i$ always accepts a finite set). We use $SUV_i$, called the surviving set, to denote all the successful test sequences. If the surviving set is empty, then the global testing problem returns "yes" (i.e., none of observable behaviors of the system $Sys$ is in $Bad$). Otherwise, if $i < k$ (i.e., it is not the last step), we goto the following **Step** $i + 1$. If $i = k$ (i.e., it is the last step and the surviving set is not empty), then the global testing problem returns "no" (i.e., some observable behaviors of the system $Sys$ is indeed in $Bad$).

In the rest of this section, we will clarify how Automata Generation and Surviving Set Generation in the $k$ steps can be done. Since our technique heavily depends on automata theory, we would like to first build the theory foundation of our technique before we proceed further.

### 4.1 Theory Foundation of the Push-in Technique

Let us first make a pessimistic (the name is borrowed from the discussions in [9]) modification of the original system $Sys$ by assuming that each black-box $B_i$, $1 \leq i \leq k$, can demonstrate *any* observable behavior in $\Sigma_i^*$ (recalling that $\Sigma_i$ is the interface of the black-box). The resulting system is denoted by $\hat{Sys}$. Clearly, every observable behavior of $Sys$ is also an observable behavior of $\hat{Sys}$ (but the reverse is not necessarily true).

Notice that $\hat{Sys}$ does not have any black-boxes since the original black-box $B_i$, after the pessimistic modification, can be considered as a finite state unit $\hat{B}_i$ with only one state, where each action in $\Sigma_i \cup \{\epsilon\}$ is a label on a transition from the state back to the state. According to the semantics definition presented in Section 3.2, it is not hard to see that $\hat{Sys}$ itself, after the composition of the gluer $G$ with all the one-state units $\hat{B}_1$, ..., $\hat{B}_k$, is a finite state transition system with $|G|$ (the number of states in the gluer) states and with actions in $\Sigma \cup \{\epsilon\}$. (Recall that $\Sigma = \Sigma_0 \cup \ldots \cup \Sigma_k$ is the union of all observable actions in the gluer and the black-boxes.) The pessimistic system can also be treated as a pessimistic (finite) automaton by making each state be an accepting state and each $\epsilon$-transition be an $\epsilon$-move. In this way, the language (a subset of $\Sigma^*$) accepted by the automaton is exactly all the observable behaviors of the pessimistic system.

As we have mentioned earlier, the set $Bad \subseteq \Sigma^*$ is a finite and hence regular set. Suppose that the symbolic representation of the set is given as an automaton $M_{Bad}$ (whose state number is written $|M_{Bad}|$); i.e., the language accepted by $M_{Bad}$ is exactly the set $Bad$.

Using a standard Cartesian product construction, one can build an automaton $M_{global}$, called the global test sequence automaton, to accept the intersection of the language accepted by the pessimistic automaton $\hat{Sys}$ and the language accepted by the automaton $M_{Bad}$. That is, $M_{global}$ accepts exactly the bad and observable behaviors of the pessimistic system. Clearly, the state number in $M_{global}$ is at most $|G| \cdot |M_{Bad}|$.

For a word $\alpha \in \Sigma^*$, we use $\alpha \downarrow_{\Sigma_i}$, $1 \leq i \leq k$, to denote the result of dropping all symbols not in $\Sigma_i$ from $\alpha$. That is, if $\alpha$ is an observable behavior of the system $Sys$, then $\alpha \downarrow_{\Sigma_i}$ is the corresponding observable behavior of black-box $B_i$. The theory foundation of our push-in technique can be summarized in the following theorem, which can be shown using the semantics defined in Section 3.2.

**Theorem 1.** *For any global test sequence $\alpha$ in $\Sigma^*$, the following two items are equivalent:*

*(1) $\alpha$ is a bad (i.e., in $Bad$) observable behavior of the system $Sys$ of black-boxes $B_1$, ..., $B_k$,*
*(2) $\alpha$ is accepted by the global test sequence automaton $M_{global}$, and each of the following $k$ conditions holds:*
  *(2.1) $\alpha \downarrow_{\Sigma_1}$ is an observable behavior of $B_1$,*
  $\vdots$
  *(2.k) $\alpha \downarrow_{\Sigma_k}$ is an observable behavior of $B_k$.*

We use "class C" to denote all the $\alpha$'s that satisfy Theorem 1 (2). Obviously, the global testing problem (i.e., there is no bad behavior in $Sys$) is equivalent to the emptiness of class C.

In the push-in technique, the jobs of **Step 1**, ..., **Step $k$** are to establish the emptiness of class C using both automata theory and black-box testing. One naive approach for the emptiness is to use Theorem 1 (2) directly: repeatedly pick a global test sequence $\alpha$ accepted by $M_{global}$ (note that $M_{global}$ accepts a finite language) and, using black-box testing, make sure that one of the conditions (2.$i$), $1 \leq i \leq k$, is false. This naive approach works but inefficiently. This is because, when $M_{global}$ accepts a huge set (such as more than $10^{24}$ in our experiments shown later), trying every such element is not only infeasible but also unnecessary. Our approach of doing the job aims at eliminating the inefficiency. First, we do not pick a global test sequence $\alpha$. Instead, we *compute* the test sequences run on black-box $B_i$ from the testing *results* on black-box $B_{i-1}$ in the previous **Step** $i - 1$. As we have mentioned at the beginning of this section, each **Step** $i$ has two tasks to perform: Automata Generation and Surviving Set Generation, which are presented in detail as follows.

### 4.2 Automata Generation in Step $i$

This task in **Step** $i$ is to generate two automata: the unit test sequence automaton $U_i$ and the auxiliary automaton $A_i$.

Initially when $i = 1$, $A_1$ is constructed as $A_1 = M_{global} \downarrow_{\Sigma_1 \cup ... \cup \Sigma_k}$, i.e., the result of dropping every transition in $M_{global}$ that is labeled with an observable action not in $\Sigma_1 \cup ... \cup \Sigma_k$. $U_1$ is constructed as the automaton $U_1 = A_1 \downarrow_{\Sigma_1}$ (i.e., the result of dropping every transition in $A_1$ that is labeled with an observable action not in $\Sigma_1$). Observe that $A_1$ accepts the language $\mathcal{A}_1 = \{\alpha \downarrow_{\Sigma_1 \cup ... \cup \Sigma_k}: \alpha \text{ accepted by } M_{global}\}$ and $U_1$ accepts the language $\mathcal{U}_1 = \{\alpha \downarrow_{\Sigma_1}: \alpha \text{ is in } \mathcal{A}_1\}$. The state number in either of the two automata, in worst cases, is $|M_{global}|$.

When $i > 1$, the two automata $A_i$ and $U_i$ are constructed from the auxiliary automaton $A_{i-1}$ and the surviving set $SUV_{i-1}$ obtained in the previous step. To construct $A_i$, we first build an automaton $suv_{i-1}$ to accept the finite set $SUV_{i-1}$. Then, we build an intermediate automaton $M_{i-1}$ that works as follows: on an input word in $(\Sigma_{i-1} \cup ... \Sigma_k)^*$, $M_{i-1}$ starts simulating $A_{i-1}$ and $suv_{i-1}$ on the word, in parallel. During the simulation, whenever $suv_{i-1}$ reads an input symbol that is not in $\Sigma_{i-1}$ (note that $suv_{i-1}$ only accepts words in $\Sigma_{i-1}^*$), it skips the input symbol. $M_{i-1}$ accepts the input word when both $A_{i-1}$ and $suv_{i-1}$ accept. Finally, the auxiliary automaton $A_i$ is constructed as $A_i = M_i \downarrow_{\Sigma_i \cup ... \Sigma_k}$. The unit test sequence automaton $U_i$ is constructed as $U_i = A_i \downarrow_{\Sigma_i}$.

One can show that each of the two automata $A_i$ and $U_i$ has, in worst cases, a state number of $|A_{i-1}| \cdot |suv_{i-1}|$. Also, $A_i$ accepts the language $\mathcal{A}_i = \{\alpha \downarrow_{\Sigma_i \cup ... \cup \Sigma_k}: \alpha \in (\Sigma_{i-1} \cup ... \cup \Sigma_k)^* \text{ is in } \mathcal{A}_{i-1} \text{ and } \alpha \downarrow_{\Sigma_{i-1}} \text{ is in } SUV_{i-1}\}$ and $U_i$ accepts the language $\mathcal{U}_i = \{\alpha \downarrow_{\Sigma_i}: \alpha \in (\Sigma_i \cup ... \Sigma_k)^* \text{ is in } \mathcal{A}_i\}$.

As we have mentioned earlier, when the empty string is accepted by the auxiliary automaton $A_i$ (a standard membership algorithm can be used to validate the acceptance), our push-in technique will return a "no" answer on the global testing problem (i.e., the system does have a bad observable behavior) and no further steps need to run.

### 4.3 Surviving Set Generation in Step $i$

The surviving set $SUV_i$ is the set of all successful unit test sequences $\alpha \in \mathcal{U}_i$; i.e., $SUV_i = \{\alpha \in \Sigma_i^* : \alpha \in \mathcal{U}_i \text{ and } \alpha \text{ is an observable behavior of black-box } B_i\}$.

A straightforward way to obtain the set is to run the black-box testing procedure **BBtest** over the black-box $B_i$ with every test sequence in $\mathcal{U}_i$. This is, however, not efficient, in particular when the set $\mathcal{U}_i$ is huge. Observable behaviors of a unit are prefix-closed: if $\alpha$ is not an observable behavior of $B_i$, then, for any $\beta$, $\alpha\beta$ can not be (i.e., test sequence $\alpha\beta$ need not be run). With prefix-closeness and **BBtest**, we use the following automata-theoretic procedure to generate the surviving set $SUV_i$.

Recall that $\mathcal{U}_i$ is a finite set of unit test sequences and, as a regular language, accepted by the unit test sequence automaton $U_i$. Let $m$ be the maximal length of all test sequences in $\mathcal{U}_i$ (the length can be obtained using a standard longest path algorithm over the transition graph of automaton $U_i$). Our procedure consists of the following $m$ jobs. Each $Job_j$, where $j$ is from 1 to $m$, is to identify (using black-box testing) all the successful test sequences (with length $j$) which are prefixes (which are not necessarily proper) of some test sequences in $\mathcal{U}_i$. In order to do this efficiently, the job makes use of the previous testing results in $\Theta_{j-1}$. More precisely, each $Job_j$ has two parts (by assumption, let $\Theta_0$ contain only the empty word.):

- Define $P_j$ to be the set of all the prefixes with length $j$ of all the unit test sequences in $\mathcal{U}_i$. Calculate the set $\hat{P}_j \subseteq P_j$ such that each element in $\hat{P}_j$ has a prefix (with length $j-1$) in $\Theta_{j-1}$. To implement this part, one can first construct an automaton (from automaton $U_i$) to accept the language $P_j$. Then, construct another automaton to accept the set $\Theta_{j-1}$. Finally, an automaton $M$ can be constructed from these two automata to accept the language $\hat{P}_j$. All the constructions are not difficult and do not involve testing.
- Using **BBtest**, generate the set $\Theta_j$ that consists of all the successful test sequences over black-box $B_i$ in $\hat{P}_j$. Hence, one only runs test sequences in $\hat{P}_j$ instead of the entire $P_j$, thanks to the previous testing results in $\Theta_{j-1}$.

It is left to the reader to verify that, after the jobs are completed, the surviving set $SUV_i$ can be obtained as $\mathcal{U}_i \cap (\cup_{0 \leq j \leq m} \Theta_j)$. Again, this set can be accepted by an automaton, treated as a symbolic representation of the set, constructed from automaton $U_i$ and the automata built in the above jobs to accept $\Theta_j$, $1 \leq j \leq m$. One can choose the procedure to output the explicit set $SUV_i$ or its symbolic representation $suv_i$.

### 4.4 Correctness and Bad Behavior Generation

Since the global testing problem is equivalent to the emptiness of class C, we only need to show that the emptiness is answered correctly with the push-in technique. Clearly, the technique always terminates with a yes/no answer. It returns "yes" only at some **Step** $i$, $1 \leq i \leq k$, whose surviving set $SUV_i = \emptyset$. It returns "no" only

CASE1. at some **Step** $i$, $1 \leq i \leq k$, when the auxiliary automaton $A_i$ accepts the empty word, or
CASE2. at the last **Step** $k$ when $SUV_k \neq \emptyset$.

In these two cases, in order to demonstrate a global bad behavior of the system, we first define an operation called $\text{select}_j(\cdot)$, $1 \leq j \leq k$. Given a sequence $\alpha_j$, the operation returns a sequence $\alpha_{j-1}$ (when $j=1$, it simply returns $\alpha_j$) satisfying the following conditions: $\alpha_{j-1} \in \mathcal{A}_{j-1}$, $\alpha_{j-1} \downarrow_{\Sigma_{j-1}} \in SUV_{j-1}$ and $\alpha_{j-1} \downarrow_{\Sigma_j \cup \ldots \Sigma_k} = \alpha_j$. The returned sequence $\alpha_{j-1}$ may not be unique. In this case, any sequence (such as a shortest one) satisfying the conditions will be fine. Now, we define another operation called $\textbf{BadGen}_j(\cdot)$, $1 \leq j \leq k$, as follows. Given a sequence $\alpha_j$, we first calculate $\alpha_{j-1} = \text{select}_j(\alpha_j)$. Then, we calculate $\alpha_{j-2} = \text{select}_{j-1}(\alpha_{j-1})$, and so on. Finally, we obtain $\alpha_1$. At this time, the operation $\textbf{BadGen}_j(\alpha_j)$ returns any sequence $\alpha$ satisfying the following conditions: $\alpha$ is accepted by $M_{global}$ and $\alpha \downarrow_{\Sigma_1 \cup \ldots \Sigma_k} = \alpha_1$. All these operations can be easily implemented through automata constructions.

Coming back to bad behavior generation, in CASE1, we return $\textbf{BadGen}_i(\lambda)$ (where $\lambda$ is the empty sequence) as a global bad behavior. In CASE2, we simply pick any sequence $\alpha_k$ from $SUV_k$ and return $\textbf{BadGen}_k(\alpha_k)$ as a global bad behavior.

One can show that our technique is indeed correct:

**Theorem 2.** *If the class C is empty then the push-in technique returns "yes", otherwise it returns "no". When the technique returns yes, it shows that the system doesn't have any of the global bad behaviors in* BAD*, otherwise it indicates that the system does exhibit bad behaviors in* BAD.

In each step of our algorithm, one can use standard algorithms in automata theory to make the obtained automata like $U_i$'s and $A_i$'s smaller. The algorithms include eliminating unreachable states and/or minimization. Additionally, the algorithms as well as all the automata constructions mentioned in the push-in technique can be implemented using existing automata manipulation tools like Grail [1].

From the correctness theorem, we know that the push-in technique is sound and complete. However, one question still remains unsolved: Are test sequences (for black-box $B_i$) in each $\mathcal{U}_i$ more than necessary (in solving the global testing problem)? We can show that each $\mathcal{U}_i$ derived from our push-in technique is "optimal" in the following sense. Suppose that we have completed the first $i-1$ **Step**s (i.e., the black-boxes $B_1, \ldots, B_{i-1}$ have been tested) and have obtained $\mathcal{U}_i$ to start the subsequent steps (i.e., the remaining black-boxes $B_i, \ldots, B_k$ are not tested yet). Each test sequence $\alpha_i$ in $\mathcal{U}_i$ has to be run, since one can show the following two statements: There are black-boxes $B_i^*, \ldots, B_k^*$, such that $\alpha_i$ is a successful (resp. unsuccessful) test sequence for $B_i^*$ and the system $G(B_1, \ldots, B_{i-1}, B_i^*, \ldots, B_k^*)$ has (resp. does not have) a global bad behavior.

|        |           | maxlength=10 | | | | maxlength=20 | | | | maxlength=30 | | | |
|--------|-----------|--------------|--|--|--|--------------|--|--|--|--------------|--|--|--|
|        | $step_i$  | $\#A_i$ | $\#U_i$ | $\#SUV_i$ | $TC_i$ | $\#A_i$ | $\#U_i$ | $\#SUV_i$ | $TC_i$ | $\#A_i$ | $\#U_i$ | $\#SUV_i$ | $TC_i$ |
| case 1 | $step_1$  | $1.06 \times 10^7$ | 148 | 47 | 68 | $7.16 \times 10^{15}$ | $8.06 \times 10^4$ | 3533 | 4572 | $2.16 \times 10^{24}$ | $4.14 \times 10^7$ | $2.23 \times 10^5$ | $2.87 \times 10^5$ |
|        | $step_2$  | $3.05 \times 10^6$ | 548 | 12 | 41 | $6.92 \times 10^{14}$ | $4.62 \times 10^5$ | 177 | 393 | $1.13 \times 10^{23}$ | $2.43 \times 10^8$ | 1331 | 2940 |
|        | $step_3$  | $4.78 \times 10^4$ | $4.78 \times 10^4$ | 7 | 39 | $1.15 \times 10^{12}$ | $1.15 \times 10^{12}$ | 58 | 297 | $1.81 \times 10^{19}$ | $1.81 \times 10^{19}$ | 274 | 1577 |
| case 2 | $step_1$  | $1.38 \times 10^7$ | 386 | 73 | 121 | $5.90 \times 10^{15}$ | $2.61 \times 10^5$ | 6697 | 9384 | $1.59 \times 10^{24}$ | $1.42 \times 10^8$ | $4.74 \times 10^5$ | $6.30 \times 10^5$ |
|        | $step_2$  | $3.12 \times 10^6$ | 142 | 13 | 25 | $4.94 \times 10^{14}$ | $5.91 \times 10^4$ | 93 | 203 | $6.99 \times 10^{22}$ | $2.53 \times 10^7$ | 645 | 1356 |
|        | $step_3$  | $7.25 \times 10^5$ | $7.25 \times 10^5$ | 0 | 47 | $1.11 \times 10^{13}$ | $1.11 \times 10^{13}$ | 0 | 277 | $1.48 \times 10^{20}$ | $1.48 \times 10^{20}$ | 0 | 1259 |
| case 3 | $step_1$  | $1.38 \times 10^7$ | 386 | 73 | 121 | $5.90 \times 10^{15}$ | $2.61 \times 10^5$ | 6697 | 9384 | $1.59 \times 10^{24}$ | $1.42 \times 10^8$ | $4.74 \times 10^5$ | $6.30 \times 10^5$ |
|        | $step_2$  | $3.12 \times 10^6$ | 142 | 13 | 25 | $4.94 \times 10^{14}$ | $5.91 \times 10^4$ | 93 | 203 | $6.99 \times 10^{22}$ | $2.53 \times 10^7$ | 645 | 1356 |
|        | $step_3$  | $7.25 \times 10^5$ | $7.25 \times 10^5$ | 0 | 47 | $1.11 \times 10^{13}$ | $1.11 \times 10^{13}$ | 13 | 359 | $1.48 \times 10^{20}$ | $1.48 \times 10^{20}$ | 129 | 2577 |
| case 4 | $step_1$  | $1.30 \times 10^6$ | 178 | 32 | 76 | $3.51 \times 10^{15}$ | $2.20 \times 10^5$ | 5507 | 8197 | $1.65 \times 10^{24}$ | $1.36 \times 10^8$ | $4.44 \times 10^5$ | $6.00 \times 10^5$ |
|        | $step_2$  | $1.02 \times 10^5$ | 97 | 0 | 14 | $9.54 \times 10^{13}$ | $1.70 \times 10^5$ | 0 | 128 | $2.39 \times 10^{22}$ | $1.22 \times 10^8$ | 0 | 906 |
|        | $step_3$  | 0 | 0 | 0 | 0 | 0 | 0 | 0 | 0 | 0 | 0 | 0 | 0 |

**Table 1.** Experiment Results: Counts of Test Sequences

## 5 Experiments

All the experiments were performed on a PC with a 800MHz Pentium III CPU and 128MB memory. The Grail [1] tool was used to perform almost all the automata operations[3]. The entire experiment process was driven by a Perl script and carried out automatically. Our experiments were run on the system of black-boxes shown in Figure 1. In the experiments, we designated black-boxes Timer, Sensor and Comm as $B_1$, $B_2$, and $B_3$, respectively. The internal implementations of the black-boxes are shown in Figures 3, 4 and 5, on which the unit testing in the experiments was performed. We have totally run twelve experiments (each experiment is a complete execution of the push-in technique), which are divided into four cases. Each of the four cases consists of three experiments, which are illustrated in detail as follows.

**Case 1** Firstly, we wish that whenever a *pause* event takes place, there should be no more *send* until a *resume* occurs. The corresponding bad behaviors are specified as a regular expression, $\Sigma^* p (\Sigma - \{r\})^* s \Sigma^*$, where $\Sigma$ is the set of all the twelve events in the system; $p$, $r$, and $s$ stand for the *pause*, *send*, and *resume*, respectively (such abbreviation will be used throughout this section). For the first experiment run in this case, we chose the $Bad$ to be all words in the regular expression that are not longer than 10 (denoted by "maxlength=10"). The remaining two experiments were run with "maxlength=20" and "maxlength=30", respectively. To understand the results shown in Table 4.4, we go through the third experiment (i.e., "maxlength=30"). The results of the experiment are shown in the box at the right upper corner in the table (i.e., under the four columns associated with "maxlength=30" and in the three rows ("step$_1$", "step$_2$", "step$_3$") associated with "case 1"). The three steps in the experiment correspond to the three **Step**s (since there are three black-boxes) in the push-in technique. The auxiliary automaton $A_1$ calculated in **Step 1** accepts totally $\#A_1 = 2.16 \times 10^{24}$ test sequences. The unit test sequence automaton $U_1$ accepts $\#U_1 = 4.14 \times 10^7$ test sequences. Using the black-box testing procedure in Section 4.3, we actually only performed $TC_1 = 2.87 \times 10^5$ unit tests over $B_1$ (the Timer), among which $\#SUV_1 = 2.23 \times 10^5$ tests survived. In **Step 2** and **Step 3**, we obtained $\#A_2, \#U_2, \#A_3, \#U_3$ similarly as shown in the table. In particular, we actually performed $TC_2 = 2940$ unit tests over the Sensor in **Step 2** and $TC_3 = 1577$ unit tests over the Comm in **Step 3**. Since the last surviving set $SUV_3$ is not empty ($\#SUV_3 = 274$), the experiment detects a global bad behavior specified in this case.

---

[3] We implemented (in C) three additional operations to manipulate automata with $\epsilon$-moves and to count the number of words in a finite language accepted by an automaton, which are not provided in Grail.

Notice that the total number of unit tests run in this experiment is $TC_1 + TC_2 + TC_3$, which is not more than $2.92 \times 10^5$. This number essentially indicates the actual "cost" of the experiment in deciding whether there is a global bad behavior specified in the case and whose length is bounded by 30. This number is quite good considering the astronomical number $\#A_1 = 2.16 \times 10^{24}$ which would be the number of integration test sequences if one run integration testing, since $M_{global} = A_1$ in the system. The other two experiments ("maxlength=10" and "maxlength=20") also detected a global bad behavior and results are shown in the first three rows under "maxlength=10" and "maxlength=20" in Table 4.4 (the costs of these two experiments, which are 148 and 5262 respectively, become much smaller).

**Case 2** The detected bad behaviors are due to the concurrency nature of these black-boxes: a $fire$ was issued before the $pause$ is sent to Timer, which eventually leads to another $send$. For instance, a global bad behavior could be like the following: $fire\ data\ send\ msg\ fire\ data\ send\ cerr\ fire\ data\ pause\ send$. From this observation, we believed that the system might also have other bad behaviors: after a $cerr$ takes place, there could be another $cerr$ coming before a $resume$ occurs. Such bad behaviors are encoded by $\Sigma^* c (\Sigma - \{r\})^* c \Sigma^*$. The three experiments in this case, however, did not detect such bad behaviors (i.e., $\#SUV_3 = 0$ for all lengths, shown in the third row "step$_3$" associated with "case 2" in Table 4.4).

**Case 3** Based upon the experiments in the previous case, we carefully studied the system and realized that the implementation of Comm might be wrong: after an error occurs (i.e., a $cerr$ outputs), Comm is supposed to retain its state prior to the output of the $cerr$, while it does not. After correcting this bug (by making the internal implementation of Comm, shown in Figure 5, move to state $s2$ instead of $s0$ after a $cerr$ is output), in this case, we run the three experiments again. The experiments detected bad behaviors only with length more than 10 (i.e., $\#SUV_3 = 0$ when maxlength is 10 and $\#SUV_3 > 0$ when maxlength is 20 and 30, shown in Table 4.4).

**Case 4** Now we want to test that: after an error occurs in Sensor (i.e., a $serr$ is issued), there will be at most one more $fire$ issued before a $resume$ occurs. The corresponding bad behaviors are encoded by $\Sigma^* serr (\Sigma - \{r\})^* f (\Sigma - \{r\})^* f (\Sigma - \{r\})^* r \Sigma^*$, where $f$ stands for $fire$. Our experiments did not detect any of such behaviors for all the three choices of maxlength: 10, 20, 30. In fact, in the experiments, no testing over Comm was needed. This is because, shown in the last three rows of Table 4.4, $\#SUV_2$ is 0 for all the three choices.

We measured the total time that our script used for automata manipulations in each of the twelve experiments, shown in Table 2. In the table, the "result" shows whether a global bad behavior was detected in an experiment; i.e., "×" (resp. "√") indicates "detected" (resp. "not detected"). As shown in the table, the total time is within a minute for all the four experiments with "maxlength=10". For "maxlength=20", the time is still acceptable (within an hour). When the maxlength is increased to 30, the time is still within our patience

(which was set to be 24 hours). Yet, our script could not finish within the patience for any experiment when we tried to push maxlength to 40. Even though determinization and minimization are optional in our push-in technique, we made them mandatory in our experiments. In this way, we can cross-compare the sizes of the automata obtained in each step of the experiments. The largest size of all the automata constructed in the twelve experiments, after determinization and minimization, is with 726 states and 2138 transitions. In an experiment with maxlength=40, the script tried to make an automaton (with 1182 states) deterministic and failed to do so within our patience.

Exhaustive integration testing over a concurrent system is in general infeasible. However, the experiments show that, using the push-in technique, we can completely solve the global testing problem with a substantially smaller number of tests over each individual black-box only, even for an extremely large set of $Bad$. For instance, the total number of unit tests ($TC_i$'s) performed in each of the four experiments with "maxlength=30" is in the order of $10^5$, while each $Bad$ is in the order of $10^{24}$ (notice that each $Bad$ is always larger than each $\#A_1$, shown in Table 4.4).

|        | maxlength=10 |        | maxlength=20 |        | maxlength=30 |        |
|--------|--------------|--------|--------------|--------|--------------|--------|
| Cases  | time         | result | time         | result | time         | result |
| Case 1 | ~25s         | ×      | ~40m         | ×      | ~19h         | ×      |
| Case 2 | ~34s         | √      | ~58m         | √      | ~18h         | √      |
| Case 3 | ~36s         | √      | ~56m         | ×      | ~18h         | ×      |
| Case 4 | ~17s         | √      | ~22m         | √      | ~5h          | √      |

**Table 2.** Experiment Results: Time Efficiency

## 6  Future Work

This paper presents an automata-theoretic and decompositional technique to testing a system of concurrent black-boxes, which is automatic, sound, and complete. Our technique can be generalized to many other forms of bad behavior specifications (i.e., the finite set $Bad$). For instance, we may that specify that $Bad$ consist of all observable sequences not longer than 40, each of which can make the gluer enter a given (undesired) state. But the exact formalisms for bad behavior specifications need further investigation. Our model of the system is based on synchronized communications. Therefore, it would be interesting to see whether the approach can be generalized to some forms of asynchronous (e.g., shared-variable) systems. Black-boxes in our model are event-driven; it is also worthwhile to study other decompositional testing approaches for data-driven black-boxes. Sometimes, our push-in technique fails to complete, due to an extremely large bad behavior set $Bad$ (e.g., our experiments with "maxlength=40" shown earlier, whose global test sequences deduced from $Bad$ are roughly as many as $10^{33}$). In this case, we need study methods to (symbolically) partition the set into smaller subsets such that the push-in technique can be run over each smaller subset. In this way, a global bad behavior could instead be found. In our definition of the push-in technique, there is not a pre-defined ordering in testing the black-boxes. For instance, in our experiments, the ordering was `Timer`, `Sensor`, `Comm`, based on the size of a black-box's interface. Clearly, more studies are needed to clarify the relationship between the efficiency of our technique and the choices of the ordering.

# References

1. Grail homepage. http://www.csd.uwo.ca/research/grail/.
2. Martn Abadi and Leslie Lamport. Composing specifications. *TOPLAS*, 15(1):73–132, 1993.
3. Rajeev Alur, Thomas A. Henzinger, Freddy Y. C. Mang, Shaz Qadeer, Sriram K. Rajamani, and Serdar Tasiran. MOCHA: Modularity in model checking. In *CAV'98*, volume 1427 of *LNCS*, pages 521–525. Springer, 1998.
4. A. Bertolino and A. Polini. A framework for component deployment testing. In *ICSE'03*, pages 221–231. IEEE Press, 2003.
5. Ed Brinksma and Jan Tretmans. Testing transition systems: An annotated bibliography. In *Proc. 4th Summer School on Modeling and Verification of Parallel Processes*, pages 187–195. Springer-Verlag, 2001.
6. J. Callahan, F. Schneider, and S. Easterbrook. Automated software testing using modelchecking. WVU Technical Report #NASA-IVV-96-022.
7. E. Clarke, D. Long, and K. McMillan. Compositional model checking. In *LICS'89*, pages 353–362. IEEE Press, 1989.
8. A. Coen-Porisini, C. Ghezzi, and R. A. Kemmerer. Specification of realtime systems using ASTRAL. *TSE*, 23(9):572–598, 1997.
9. Luca de Alfaro and Thomas A. Henzinger. Interface automata. In *FSE'01*, pages 109–120. ACM Press, 2001.
10. A. Engels, L.M.G. Feijs, and S. Mauw. Test generation for intelligent networks using model checking. In *TACAS'97*, volume 1217 of *LNCS*, pages 384–398. Springer, 1997.
11. Kathi Fisler and Shriram Krishnamurthi. Modular verification of collaboration-based software designs. In *ESEC/FSE'01*, pages 152–163. ACM Press, 2001.
12. C. Flanagan and S. Qadeer. Thread-modular model checking. In *SPIN'03*, volume 2648 of *LNCS*, pages 213–225. Springer, 2003.
13. Angelo Gargantini and Constance Heitmeyer. Using model checking to generate tests from requirements specifications. In *ESEC/FSE'99*, volume 1687 of *LNCS*, pages 146–163. Springer, 1999.
14. S. Ghosh and P. Mathur. Issues in testing distributed component-based systems. In *First ICSE Workshop on Testing Distributed Component-Based Systems*, 1999.
15. Dimitra Giannakopoulou, Corina S. Pasareanu, and Jamieson M. Cobleigh. Assume-guarantee verification of source code with design-level assumptions. In *ICSE'04*, pages 211–220. IEEE Press, 2004.
16. Mary Jean Harrold. Testing: a roadmap. In *Proceedings of the conference on the future of software engineering*, pages 61–72. ACM Press, 2000.
17. Thomas A. Henzinger, Ranjit Jhala, Rupak Majumdar, , and Shaz Qadeer. Thread-modular abstraction refinement. In *CAV'03*, volume 2725 of *LNCS*, pages 262–274. Springer, 2003.
18. Ralph D. Jeffords and Constance L. Heitmeyer. A strategy for efficiently verifying requirements. In *FSE'03*, pages 28–37. ACM Press, 2003.
19. C.B. Jones. Tentative steps towards a development method for interfering programs. *TOPLAS*, 5(4):596–619, 1983.
20. Pramod V. Koppol, Richard H. Carver, and Kuo-Chung Tai. Incremental integration testing of concurrent programs. *TSE*, 28(6):607–623, 2002.
21. Leslie Lamport. Specifying concurrent program modules. *TOPLAS*, 5(2):190–222, 1983.
22. Harry Li, Shriram Krishnamurthi, and Kathi Fisler. Verifying cross-cutting features as open systems. *ACM SIGSOFT Software Engineering Notes*, 27(6):89–98, 2002.
23. N. Lynch and M. Tuttle. An introduction to input/output automata. *CWI-Quarterly*, 2(3):219–246, 1989.
24. Lev Nachmanson, Margus Veanes, Wolfram Schulte, Nikolai Tillmann, and Wolfgang Grieskamp. Optimal strategies for testing nondeterministic systems. In *ISSTA'04*, pages 55–64. ACM Press, 2004.
25. Alessandro Orso, Mary Jean Harrold, and David S. Rosenblum. Component metadata for software engineering tasks. In *EDO'00*, volume 1999 of *LNCS*, pages 129–144. Springer, 2000.
26. Doron Peled, Moshe Y. Vardi, and Mihalis Yannakakis. Black box checking. In *FORTE/PSTV'99*, pages 225–240. Kluwer, 1999.
27. Alexandre Petrenko, Nina Yevtushenko, and Jia Le Huo. Testing transition systems with input and output testers. In *TestCom'03*, volume 2644 of *LNCS*, pages 129 – 145. Springer, 2003.

28. A. Pnueli. In transition from global to modular temporal reasoning about programs, 1985. In K.R. Apt, editor, Logics and Models of Concurrent Systems, sub-series F: Computer and System Science.
29. D. Rosenblum. Adequate testing of componentbased software. Department of Information and Computer Science, University of California, Irvine, Technical Report 97-34, August 1997.
30. C. Szyperski. Component technology: what, where, and how? In *ICSE'03*, pages 684–693. IEEE Press, 2003.
31. C. Tai and R. H. Carver. Testing of distributed programs. In *Parallel and Distributed Computing Handbook*, pages 955–978. McGraw-Hill, 1996.
32. Jan Tretmans and Ed Brinksma. Torx: Automated model-based tesing. In *First European Conference on Model-Driven Software Engineering*, pages 31–43, 2003.
33. J. Voas. Developing a usage-based software certification process. *IEEE Computer*, 33(8):32–37, August 2000.
34. John Whaley, Michael C. Martin, and Monica S. Lam. Automatic extraction of object-oriented component interfaces. In *ISSTA'02*, pages 218–228. ACM Press, 2002.
35. Gaoyan Xie and Zhe Dang. Model-checking driven black-box testing algorithms for systems with unspecified components. In *FATES'04*, LNCS, (to appear).